\documentclass[acmlarge,11pt,nonacm,screen]{acmart}
\usepackage{ifthen}
\usepackage{fancyhdr}
\usepackage{fancybox}
\usepackage{xspace}
\usepackage{booktabs}
\usepackage{mathptmx}   
\usepackage{courier}
\usepackage{color}
\usepackage{graphicx}
\usepackage{multicol}
\usepackage{multirow}
\usepackage{adjustbox}
\usepackage{verbatim}
\usepackage{appendix}
\usepackage{titlesec}

\newcommand{\TITLE}{Software Supply Chain Attribute Integrity (SCAI)}
\newcommand{\SUBTITLE}{Specification Version 0.2}

\newcommand{\PAGENUMBERS}{yes}

\newcommand{\AUTHORS}{
\author{Marcela S. Melara}
\affiliation{%
\jobtitle{Research Scientist}
  \institution{Intel Corporation}
  }
\email{marcela.melara@intel.com}
}

\newcommand{\ANONYMOUS}{no}

\newcommand{\COMMENTS}{yes}

\usepackage[absolute]{textpos}

\usepackage{caption}
\usepackage{listings}

\colorlet{numb}{magenta!60!black}
\lstdefinelanguage{json}{
    basicstyle=\normalfont\ttfamily,
    commentstyle=\color{gray}, 
    stringstyle=\color{black}, 
    numbers=none,
    numberstyle=\scriptsize,
    stepnumber=1,
    numbersep=8pt,
    showstringspaces=false,
    breaklines=true,
    frame=none,
    backgroundcolor=\color{white}, 
    string=[s]{"}{"},
    comment=[l]{//},
    morecomment=[s]{/*}{*/},
    literate=
        *{0}{{{\color{numb}0}}}{1}
         {1}{{{\color{numb}1}}}{1}
         {2}{{{\color{numb}2}}}{1}
         {3}{{{\color{numb}3}}}{1}
         {4}{{{\color{numb}4}}}{1}
         {5}{{{\color{numb}5}}}{1}
         {6}{{{\color{numb}6}}}{1}
         {7}{{{\color{numb}7}}}{1}
         {8}{{{\color{numb}8}}}{1}
         {9}{{{\color{numb}9}}}{1}
         {:}{{{\color{purple}{:}}}}{1}
         {,}{{{\color{purple}{,}}}}{1}
         {\{}{{{\color{blue}{\{}}}}{1}
         {\}}{{{\color{blue}{\}}}}}{1}
         {[}{{{\color{blue}{[}}}}{1}
         {]}{{{\color{blue}{]}}}}{1}
}

\newcommand{\eg}{e.g.,\xspace}
\newcommand{\ie}{i.e.,\xspace}

\newcommand{\Parabreak}{1.5ex}
\newcommand{\Paragraph}[1]{\vspace{\Parabreak}\noindent\textbf{#1}}
\newcommand{\swsc}{software supply chain\xspace}
\newcommand{\intelreg}{Intel\textsuperscript{\textregistered}\xspace}

\newcommand{\ignore}[1]{}

\ifthenelse{\equal{\COMMENTS}{yes}}{%
\newcommand{\msm}[1]{\textbf{MSM: #1}}
} {
\newcommand{\msm}[1]{}
}

\usepackage{enumitem}
\setlist[itemize]{noitemsep,nolistsep}
\setlist[enumerate]{noitemsep,nolistsep}

\usepackage{colortbl}
\definecolor{Gray}{gray}{0.9}
\definecolor{LightGreen}{rgb}{0.2,1,0.4}

\usepackage{longtable}

\usepackage{fancyhdr} 

\ifthenelse{\equal{\PAGENUMBERS}{yes}}{%
  \pagestyle{plain}
}{%
  \pagestyle{empty}
}

\date{}
\title{\TITLE}
\ifthenelse{\equal{\ANONYMOUS}{no}}
{\AUTHORS}{\author{}}
\subtitle{\SUBTITLE}

\setcopyright{rightsretained}
\acmDOI{}

\begin{document}
\maketitle

\ifthenelse{\equal{\PAGENUMBERS}{yes}}{%
}{\thispagestyle{empty}}

\newpage
{\hypersetup{hidelinks}
\tableofcontents
}

\newpage
\section{Introduction}
\label{sec:intro}

The Software Supply Chain Attribute Integrity, or SCAI (pronounced ``sky''), specification 
proposes a data format for capturing functional attribute and integrity information about
software artifacts and their supply chain.
SCAI data can be associated with executable binaries, statically- or dynamically-linked
libraries, software packages, container images, software toolchains, and compute
environments.

As such, SCAI is intended to be implemented as part of an existing software supply chain 
attestation framework by software development tools or services 
(\eg builders, CI/CD pipelines, software analysis tools) seeking to capture more 
granular information about the attributes and behavior of the software artifacts they produce. 
That is, SCAI assumes that implementers will have appropriate processes and tooling in place 
for capturing other types of software supply chain metadata, which can be extended to add support for
SCAI.

\subsection{Rationale}
The security of software artifacts has traditionally been established 
based on the (digital) identity of the originator and/or cryptographic hashes of the artifact. 
In the case of open-source software (OSS), trust in an artifact may also be 
based on the assumption that community members regularly monitor OSS code repositories to
detect any existing security issues. 

However, \swsc attacks such as the SolarWinds hack~\cite{solarwinds-hack}, in which the adversary
subverted the build stage of the supply chain and injected a backdoor among legitimate company code,
demonstrate that software originating from reputable sources may still not be trustworthy. In other words,
identity information is not sufficient to establish trust in software artifacts.

Existing data formats for software supply chain integrity do not fully address this issue and 
provide only coarse-grain information about a given \emph{process} in an artifact's supply chain. 
For example, Software Package Data Exchange (SPDX)~\cite{spdx} only provides a list of components (and their hashes) 
that were inputs to a build. Supply-chain Levels for Software Artifacts (SLSA) Build Provenance~\cite{slsa} 
conveys configuration and basic integrity information about the build service.

None of the existing supply chain data formats capture any information about the security \emph{functionality} or behavior
of the resulting software artifact, nor do they provide sufficient~\emph{evidence} to support any claims
of integrity of the supply chain processes they describe. The SCAI data format is designed to bridge this gap. 

\subsection{Overview}
SCAI metadata comprises a set of statements that describe functional attributes of a software artifact and its
supply chain, capable of covering the full software stack of the toolchain that produced the
artifact down to the hardware platform.
These statements, broadly referred to as Attribute Assertions (\S\ref{sec:artifact-property}), include
information about the conditions under which certain functional attributes arise, as well as (authenticated) evidence 
for the asserted attributes.
Together, this information provides a high-integrity description of the functionality of a software artifact
and the integrity of its supply chain, which enables a human or program to determine the trustworthiness of a given 
artifact based on specific attributes.

Attribute Assertions are collected during a specific operation in the software supply chain, and bound into a digitally 
verifiable unit referred to as a Report (\S\ref{sec:report}) for that operation.
Reports may be bound to the corresponding artifact, or they may be distributed through a dedicated supply chain
metadata store (\eg~\cite{sigstore}).

\subsection{SCAI Workflow: An Example}
Library developer Leo wants to create a trustworthy shared library using
a SCAI-aware compiler toolchain (known as a producer in SCAI).
To accomplish this goal, Leo builds the library using a SCAI-aware \texttt{gcc} toolchain running on a hardware-attested 
platform such as~\intelreg SGX~\cite{sgx} to reduce the risk of tampering with the compilation process.
Leo also wants to provide artifact-level security guarantees for the shared library, such as enabled buffer overflow protection, so
the toolchain is configured with the \texttt{-fstack-protector} compiler flag.

The SCAI-aware toolchain records two Assertions each describing one attribute of the artifact or its supply chain: 
(1) the ``attested platform'' attribute, which captures the given version of the gcc compiler and the associated hardware attestation as evidence, 
and (2) the ``stack protected'' attribute, in which the compiler flag \texttt{-fstack-protector} is the condition for the attribute
to apply to the artifact.

If desired, Leo's SCAI-aware toolchain may also record additional information about the compiler's invocation or
artifact components via other existing supply chain metadata formats such as SLSA~\cite{slsa} or SPDX~\cite{spdx}.
At the end of the compilation, Leo obtains a digitally signed SCAI Report containing all Attribute Assertions and extra metadata 
for the shared library.

Say that an application developer Ana wants to import Leo's shared library as a dependency
into a software package she is writing (known as a consumer in SCAI). 
Ana will first run a program to verify the digital signature of the associated SCAI Report
for the shared library against a set of trusted SCAI Report signers.

Assuming she trusts Leo's SCAI-aware compiler toolchain, Ana's verifier program 
may then validate each Attribute Assertion contained in the Report based on her trust requirements 
for shard libraries. 
In the case of Leo's library, Ana can verify that the library has the required buffer overflow protection features,
and gain confidence that the build was not compromised by verifying the hardware attestation included in the
toolchain's Attribute Assertion.
A SCAI-aware toolchain used to create Ana's software package may then record an Attribute Assertion for the package
describing a ``software-attested dependency'' that includes the SCAI Report received for Leo's library, 
thereby establishing a verifiable link between Ana's artifact and its dependencies that consumers of Ana's package may 
later transitively verify.

\subsection{SCAI Goals}
\Paragraph{Attribute-Based Trust.}
SCAI aims to enable interested parties to make informed decisions about the trustworthiness
and security of the software artifact based on concrete attributes that describe the functionality or
behavior of the artifact.

\Paragraph{Fine-Grained Provenance Rooted In Hardware.}
Many attributes or functional features of software are applied or implemented \emph{as part of} a
specific operation in its supply chain, and are influenced not only by the toolchain performing the specific
supply chain operation, but also its underlying compute environment down to the hardware platform.
As such, SCAI is capable of capturing attribute metadata at a fine granularity covering
all components involved in a given supply chain operation to obtain fine-grained provenance information
about how an artifact was created that is rooted in hardware, whenever possible.

\Paragraph{High-Integrity Metadata.}
To enable the validation of the integrity of the metadata itself,
SCAI captures evidence for claimed artifact and producer attributes. This evidence may
originate from a variety of sources including third-party auditors or trust delegates,
program analysis tools, or attested hardware, and may be represented using SCAI or other
compatible data formats. 

\Paragraph{Verifiable Dependency Graph.}
Because dependencies have a direct impact on the functionality or behavior of a software artifact,
the dependency graph of an artifact can provide useful information about its attributes and trustworthiness.
As such, SCAI can be used to explicitly bind an artifact to attribute metadata\footnote{generated with SCAI or other format-compliant specifications; see~\S\ref{sec:format-reqs}} 
for its dependencies thereby establishing a verifiable dependency graph.

\Paragraph{Application-Agnosticism.}
SCAI is agnostic to application domain, including application programming language. 
That is, SCAI does not dictate the types of artifacts, producers, supply chain toolchains, compute platforms,
attributes, evidence information, or other associated metadata that can be described with the format.
This also means that SCAI is capable of capturing non-security related artifact and
supply chain attributes. Thus, we also envision SCAI enabling other operational decisions, such as resource
selection or orchestration, based on fine-grained artifact attributes.

\Paragraph{Size-Efficiency.}
Software supply chain metadata commonly includes descriptions or pointers to 
a number of different types of artifacts or other metadata.
Thus SCAI aims to leverage new and existing supply chain metadata stores, avoid data duplication, and reduce the 
size of SCAI metadata.

\Paragraph{Interoperability.}
SCAI aims to complement existing supply chain integrity
metadata formats with attribute-based provenance information, rather than replace these formats.
SCAI interoperates with commonly used frameworks such as in-toto~\cite{in-toto}, SLSA~\cite{slsa}, SPDX~\cite{spdx},
and NIST Secure Software Development Framework (SSDF)~\cite{ssdf}.
In addition, SCAI is complementary to the C2PA~\cite{c2pa} framework, which captures provenance and authenticity
information for digital media content, such as video and digital images. 

NIST's Open Security Controls Assessment Language (OSCAL) Component Definition Model~\cite{oscal} 
enable organizations to list security features implemented by a given artifact to automate compliance checks with respect to a number of
regulatory standards. While SCAI and OSCAL Component Definitions have a similar intent, SCAI metadata could be viewed as verifiable 
evidence or proof for the features described in OSCAL Component Definition metadata for a given artifact because SCAI attributes are 
captured and digitally signed by the producer of the artifact during its creation process.

See~\S\ref{sec:examples} for examples of how SCAI may interoperate with a number of existing metadata formats.

\subsection{Scope}
This specification introduces two key data structures:
the Attribute Assertion (\S\ref{sec:artifact-property}), and the Report (\S\ref{sec:report}).
For each data structure, this document describes the requirements for creating the structure,
its fields, their purpose and intent, and their data type.

\subsection{Format Requirements}
\label{sec:format-reqs}
\begin{enumerate}
\item Must be in syntax that a software tool can read and write.
\item Syntax must be automatically verifiable for correctness, regardless of how a data object was generated (manual or tool).
\item Multiple file formats can be used to represent the captured metadata. Currently supported formats include:
\begin{itemize}
\item \textbf{JavaScript Object Notation (JSON)} see~\S\ref{sec:examples} for SCAI JSON format examples.
\end{itemize}
\end{enumerate}
\newpage
\section{Glossary}
\subsection{Artifact}
In SCAI, an ``artifact'' refers to any unit of digital information, typically represented as a file,
that is associated with the configuration, creation, distribution, or deployment of software. 
An artifact is typically produced either manually by human developers (\eg source code) or
generated automatically through software tooling (\eg compiler). 

Artifacts in SCAI include, but are not limited to, the following examples: 
\begin{itemize}
\item a source code file (\eg .c or .java)
\item a configuration file (\eg TOML or YAML)
\item a compressed archive (\eg tarball or zip)
\item an executable binary file (\eg .exe or ELF file)
\item a software package (\eg .so or .wheel)
\item a container image (\eg Docker or Gramine-SGX)
\item a manifest file (\eg for a software package or container)
\end{itemize}

SCAI is designed to be agnostic to file type or generation method, and only requires that artifacts 
be identifiable via cryptographic digests/checksums and/or a URI.

\subsection{Producer}
The ``producer'' refers to the process through which an artifact and its associated metadata is created.
In practice, a producer encompasses an artifact, as is the case with an executable compiler tool, 
or as a service, as is the case with a build or CI/CD service, and any underlying 
software, compute environment and hardware platform.

\subsection{Consumer}
The ``consumer'' refers the procedure that ingests an artifact and /or its associated
metadata either for transformation into another artifact, or for evaluation of its integrity and contents
based on a trust policy.
In practice, a consumer may represent an artifact, as is the case with an executable binary analysis tool, 
or a service, as is the case with an orchestration policy engine.

\subsection{Provenance}
In SCAI, ``provenance'' refers to information describing the producer and creation process of an artifact,
including all dependencies, and software or hardware components involved. This is not to be confused
with SLSA Provenance~\cite{slsa}, which is a data format specification for capturing metadata about a given step
in a software supply chain.

\subsection{Object}
An ``object'' in SCAI refers to any Artifact or unit of metadata. Objects must be self-describing, \ie they
must have a discreet type associated with them, such as a file type or schema.

\subsection{Reference}
A ``reference'' in SCAI refers to a data structure that describes, at the very least, 
the name, cryptographic digest, resolvable location and type of an object, enabling consumers to 
locate and validate the described object.

\subsection{Attribute}
An ``attribute'' refers to a discreet functionality, trait, or behavior found in an individual artifact. 
Typically, a producer will claim that its operation applies certain attributes to an artifact.
An attribute should be expressed in a format readable by humans and software alike.

Though SCAI does not prescribe attribute formats for SCAI metadata objects, examples of attributes that 
may be captured with SCAI include, but are not limited to:
\begin{itemize}
\item ``stack protected''
\item ``buffer overflow protected''
\item ``stack protection enabled''
\item ``stack canaries = true''
\item ``built with stack canaries''
\item WITH\_STACK\_PROTECTION
\end{itemize}

See~\S\ref{field:attr} for specification details.

\subsection{Condition}
A ``condition'' in SCAI refers to a specific configuration or state of the producer that
causes a claimed attribute to arise. Examples of a condition include, but are not limited to, invocation parameters,
configuration files, or environment variables.
See~\S\ref{field:cond} for specification details.

\subsection{Evidence}
In SCAI, ``evidence'' refers to any information that can be used to corroborate attributes
claimed to be present in an artifact. In practice, this information will often be digitally signed
in order to authenticate the originator of the evidence, and should be conveyed in an interoperable format
such as the \href{https://github.com/in-toto/attestation/blob/main/spec/v1/resource_descriptor.md}{Resource Descriptor}.

While SCAI may be used to represent evidence data, SCAI does not prescribe evidence formats. 
Examples of evidence may include, but are not limited to:
\begin{itemize}
\item an endorsement for an artifact or producer by a trusted third party
\item a log file generated by a producer
\item a report generated by a binary analysis tool
\item a vulnerability scan report
\item the results of a reproducible build~\cite{reproducible-builds}
\item a software attestation (\eg in-toto attestation~\cite{ita})
\item a hardware attestation from a trusted execution environment
\item a proof generated by a formal verification tool
\end{itemize}

See~\S\ref{field:evi} for specification details.

\subsection{Assertion}
An ``assertion'' refers to a 4-tuple data structure that describes a claimed attribute, 
a set of conditions, evidence, and a possible target object to which the claimed attribute applies.
See~\S\ref{sec:artifact-property} for specification details.
\newpage
\section{Object Reference (Deprecated)}
\label{sec:obj-ref}

A SCAI Object Reference is designed to be a size-efficient representation of any
object, artifact or metadata, that may be included in SCAI metadata. The Object Reference must 
allow both humans and automated 
verifier programs to easily parse, identify and locate the referenced objects.

\Paragraph{Since v0.2:} For consistency and improved interoperability between
producers and consumers of SCAI metadata, the SCAI Object Reference introduced in 
\href{https://arxiv.org/pdf/2210.05813v1.pdf}{v0.1} has been subsumed by
the
\href{https://github.com/in-toto/attestation/blob/main/spec/v1/resource_descriptor.md}{Resource Descriptor}
format. For the full specification and requirements, please refer to the latest Resource Descriptor
documentation in the in-toto Attestation Framework~\cite{ita}.
The Object Reference specification will be removed in a future version of the SCAI specification.

\subsection{Object Type}

\Paragraph{Purpose:} Optional field indicating the type of referenced object.

\Paragraph{Intent:} When included, the Object Type enables the consumer
to validate type-specific features. Typically, the Object Type for an artifact will be its file type.
The Object Type for a metadata object will commonly be a data format or schema identifier.

\Paragraph{Data type:} String.

\subsection{Name}

\Paragraph{Purpose:} Human-readable identifier to distinguish the referenced object.

\Paragraph{Intent:} The Name enables both humans and machines to quickly match objects.
The semantics are up to the producer and consumer.

\Paragraph{Data type:} String.

\subsection{Digest}

\Paragraph{Purpose:} A set of cryptographic digests of the referenced object.

\Paragraph{Intent:} The Digest enables a consumer to match objects and validate their integrity.
The producer and consumer must agree on acceptable algorithms. If there are no
overlapping algorithms, the subject is considered not matching.

\Paragraph{Data type:} Object following the 
\href{https://github.com/in-toto/attestation/blob/v0.1.0/spec/field_types.md#DigestSet}{DigestSet} format.

\subsection{Location URI}

\Paragraph{Purpose:} The location of the referenced object.

\Paragraph{Intent:} The producer indicates the Location URI to enable a consumer to automatically 
locate and obtain the referenced object.
Acceptable locations (web server, local, git etc.) are up to the producer
and consumer. To enable automated downloads, the Location URI SHOULD be resolvable.

\Paragraph{Data type:} Universal Resource Identifier (URI) string as specified in \href{https://www.rfc-editor.org/rfc/rfc3986}{RFC 3986}.

\newpage
\subsection{Format in JSON}
\begin{lstlisting}[language=json, firstnumber=1]
{
    "objectType": "<TYPE>", // optional
    "name": "<NAME>",
    "digest": { "<ALGORITHM>": "VALUE", "<ALGORITHM>": "VALUE", .. },
    "locationURI": "<URI>"
}
\end{lstlisting}

\newpage
\section{Attribute Assertion}
\label{sec:artifact-property}

Attribute Assertions must allow humans and programs
to easily parse the asserted attributes. Additional 
fields must enable program-based consumers to automatically parse and evaluate
the given information.

\subsection{Attribute}
\label{field:attr}

\Paragraph{Purpose:} An attribute of an artifact or artifact producer.

\Paragraph{Intent:} The Attribute indicates a specific functional feature of the
artifact or artifact producer is claimed to have. Attributes are expected to be domain- or application-specific.
Acceptable Attribute descriptors and formats are up to the producer and consumer.

\Paragraph{Data type:} String.

\subsection{Conditions}
\label{field:cond}
\Paragraph{Purpose:} Optional field indicating conditions for the Attribute.

\Paragraph{Intent:} The producer may use Conditions to describe specific conditions under which
the asserted Attribute hold true. Acceptable Conditions formats are up to the 
producer and consumer. 

\Paragraph{Data type:} Object following the SCAI format requirements (\S\ref{sec:format-reqs}). 
Regardless of format, the Conditions object MUST be self-describing (\eg have a type URI) 
to enable the consumer to evaluate the object against a policy.

\subsection{Target}

\Paragraph{Purpose:} Optional field indicating a target object for the Attribute.

\Paragraph{Intent:} The producer may use Target to provide a reference to a specific object (artifact or metadata) to which the
information specified in the Attribute (\S\ref{field:attr}) field applies. 

\Paragraph{Data type:} Object following the \href{https://github.com/in-toto/attestation/blob/main/spec/v1/resource_descriptor.md}{Resource Descriptor} format.

\subsection{Evidence}
\label{field:evi}
\Paragraph{Purpose:} Optional field indicating any evidence information for the Attribute.

\Paragraph{Intent:} The producer may use Evidence to provide a reference to information that serves
as evidence for the asserted Attribute. Acceptable Evidence descriptions are up to the 
producer and consumer. 

\Paragraph{Data type:} Object following the \href{https://github.com/in-toto/attestation/blob/main/spec/v1/resource_descriptor.md}{Resource Descriptor} format.

\subsection{Format in JSON}
\begin{lstlisting}[language=json, firstnumber=1]
{
    "attribute": "<ATTRIBUTE>",
    "conditions": { /*object */ }, // optional
    "target": { /* Resource Descriptor */  } // optional
    "evidence": { /* Resource Descriptor */ }, // optional
}
\end{lstlisting}

\newpage
\section{Report}
\label{sec:report}

A SCAI Report describes a set of Attribute Assertions(\S\ref{sec:artifact-property}) 
about a subject artifact and its dependencies. When applicable, SCAI Reports also contain information about the 
producer, including Attribute Assertions about its compute environment 
and metadata about its operation.

\Paragraph{Recommendation:} To aid interoperability and consistent formatting of SCAI Reports, the
in-toto compatible \href{https://github.com/in-toto/attestation/blob/main/spec/predicates/scai.md}{SCAI predicate} 
should be used within the larger in-toto Attestation Framework~\cite{ita}, which provides a standard format, and 
further options and requirements, for the Subject and Signature fields described in this specification.

\subsection{Subject}
\label{field:subject}

\Paragraph{Purpose:} The artifact the Report refers to.

\Paragraph{Intent:} The Subject indicates the object (typically an artifact) to which the Report applies. 

\Paragraph{Data type:} Object following the \href{https://github.com/in-toto/attestation/blob/main/spec/v1/resource_descriptor.md}{Resource Descriptor} format.
As Subjects should point to immutable artifacts, at least the digest field of the Resource Descriptor is required. This is aligned with
the current in-toto specification~\cite{ita} for \href{https://github.com/in-toto/attestation/blob/main/spec/v1/statement.md#fields}{attestation subjects}.

\subsection{Attributes}
\label{field:attrs}
\Paragraph{Purpose:} A set of attributes describing the Subject (\S\ref{field:subject})
artifact.

\Paragraph{Intent:} The Attributes enable a producer to make assertions about specific
attributes of a software artifact. The semantics of the specified attributes are up to the producer and consumer.

\Paragraph{Data type:} List of one or more objects following the Attribute Assertion (\S\ref{sec:artifact-property}) format.

\Paragraph{Since v0.2:} For added simplicity of the Report structure, the Subject Attributes field
introduced in \href{https://arxiv.org/pdf/2210.05813v1.pdf}{v0.1} has been renamed to Attributes.

\subsection{Producer}
\Paragraph{Purpose:} Optional field describing the the producer of the Subject (\S\ref{field:subject}) artifact and Attributes (\S\ref{field:attrs}).

\Paragraph{Intent:} The Producer field can be used to provide information about the producer, including its compute environment and any relevant metadata.

\Paragraph{Data type:} Object following the \href{https://github.com/in-toto/attestation/blob/main/spec/v1/resource_descriptor.md}{Resource Descriptor} format.

\Paragraph{Since v0.2:} For added simplicity of the Report structure, the Attributes and Operation subfields
of the Producer have been removed, and replaced with a single, more compact 
\href{https://github.com/in-toto/attestation/blob/main/spec/v1/resource_descriptor.md}{Resource Descriptor} format
field. We recommend that assertions about producers be captured in separate Reports with the intended
producer as the Subject (\S\ref{field:subject}).

\subsection{Signature}
\Paragraph{Purpose:} The digital signature over the contents of the Report.

\Paragraph{Intent:} The Signature authenticates the Report and is computed over the
Subject, Attributes, and Producer fields. Acceptable representation formats
for the Signature are up to the producer and consumer. 

\Paragraph{Data type:} Object following the SCAI format requirements (\S\ref{sec:format-reqs}). 
Regardless of format, the Signature object MUST be self-describing to enable the 
consumer to evaluate the object against a policy.

\subsection{Format in JSON}
\begin{lstlisting}[language=json, firstnumber=1]
{
    "subject": { 
        "name": "<NAME>", // optional
        "uri": "<URI>" // optional
        "digest": { "<ALGORITHM>": "VALUE" }, // required
        ...
    }
    "attributes": [{
        "attribute": "<ATTRIBUTE>",
        "target": { /* Resource Descriptor */ },
        "conditions": { /* object */ },
        "evidence": { /* Resource Descriptor */ }
    }, {...} ],
    "producer": { // optional
        "name": "<NAME>", // optional
        "uri": "<URI>",
        ...
    },
    "signature": { /* object */ }
}
\end{lstlisting}
\newpage
\section{Metadata Examples}
\label{sec:examples}

All examples for SCAI are shown in JSON format. We include examples
of in-toto Resource Descriptors~\cite{ita} for added clarity and completeness.

For examples of SCAI usage with the in-toto Attestation Framework~\cite{ita},
please refer to the 
\href{https://github.com/in-toto/attestation/blob/main/spec/predicates/scai.md#examples}{SCAI predicate examples} page.

\subsection{Resource Descriptor for source file}
\begin{lstlisting}[language=json, firstnumber=1]
{
    "name": "hello-world.c", 
    "digest": { "sha256": "e5564d4..." },
    "downloadLocation": "http://example.com/sources/hello-world.c",
}
\end{lstlisting}

\subsection{Resource Descriptor for software package}
\begin{lstlisting}[language=json, firstnumber=1]
{
    "name": "gcc9.3.0", 
    "digest": { "sha256": "78ab6a8..." },
    "downloadLocation": "http://us.archive.ubuntu.com/ubuntu/pool/main/g/gcc-defaults/gcc_9.3.0-1ubuntu2_amd64.deb",
    "mediaType": "application/vnd.debian"
}
\end{lstlisting}

\subsection{Attribute Assertion for compiler flags}
\begin{lstlisting}[language=json, firstnumber=1]
{
    "attribute": "WITH_STACK_PROTECTION", 
    "conditions": { "flags": "-fstack-protector*" }
}
\end{lstlisting}

\subsection{Attribute Assertion for compiler flags with endorsement}
\begin{lstlisting}[language=json, firstnumber=1]
{
    "attribute": "WITH_STACK_PROTECTION", 
    "conditions": { "flags": "-fstack-protector*" }
    "evidence": {
        "name": "gcc9.3.0-endorsement", 
        "digest": { "sha256": "abababa..." },
        "uri": "http://example.com/scai-reports/gcc9.3.0-endorsement",
        "mediaType": "application/json"
    }
}
\end{lstlisting}

\newpage
\subsection{Attribute Assertion for \intelreg SGX trusted execution~\cite{sgx}}
\begin{lstlisting}[language=json, firstnumber=1]
{
    "attribute": "ATTESTED_HARDWARE", 
    "target": {
        "name": "enclave.signed.so", 
        "digest": { "sha256": "e3b0c44..." },
        "downloadLocation": "http://example.com/enclaves/enclave.signed.so",
    },
    "evidence": {
        "digest": { "sha256": "0987654..." },
        "downloadLocation": "http://example.com/sgx-attestations/my-sgx-builder.json",
        "mediaType": "application/x.sgx.dcap1.14+json"
   }
}
\end{lstlisting}

\subsection{Report for basic gcc compiler endorsement with JWS signature~\cite{jws}}
\begin{lstlisting}[language=json, firstnumber=1]
{
    "subject": {
        "name": "gcc9.3.0", 
        "digest": { "sha256": "78ab6a8..." },
        "downloadLocation": "http://us.archive.ubuntu.com/ubuntu/pool/main/g/gcc-defaults/gcc_9.3.0-1ubuntu2_amd64.deb",
        "mediaType": "application/vnd.debian"
    },
    "attributes": [{
        "attribute": "WITH_STACK_PROTECTION", 
        "conditions": { "flags": "-fstack-protector*" }
    },
    {
        "attribute": "ENDORSED"
    }],
    "signature": {
        "type": "https://datatracker.ietf.org/doc/rfc7515/",
        // JWS signature (flattetend)
        "payload": "eyJfdHlwZSI6ICJodHRwczovL2lu...",
        "protected":"kjbrfkjekrf...",
        "header": { "kid":"deadead..." },
        "signature": "DGkldfhHDG..."
    }
}
\end{lstlisting}

\newpage
\subsection{Report for reproducible gcc build, with SLSA Provenance~\cite{slsa}, DSSE signature~\cite{dsse}}
\begin{lstlisting}[language=json, firstnumber=1]
{
    "subject": {
        "name": "gcc9.3.0", 
        "digest": { "sha256": "78ab6a8..." },
        "downloadLocation": "http://us.archive.ubuntu.com/ubuntu/pool/main/g/gcc-defaults/gcc_9.3.0-1ubuntu2_amd64.deb",
        "mediaType": "application/vnd.debian"
    },
    "attributes": [{
        "attribute": "REPRODUCIBLE", 
        "evidence": {
            "name": "rebuilderd-attestation", 
            "digest": { "sha256": "abcdabcde..." },
            "downloadLocation": "http://example.com/rebuilderd-instance/gcc_9.3.0-1ubuntu2_amd64.att",
            "mediaType": "https://in-toto.io/link/v0.1"   
        }
    },
    {
        "attribute": "SLSA_PROVENANCE", 
        "evidence": {
            "uri": "http://example.com/rekor-instance/slsa-provenance-gcc9.3.0",
            "digest": {"sha256": "01012424..."},
            "mediaType": "application/x.dsse+json"
       	}
    }],
    "signature": {
        "type": "https://github.com/secure-systems-lab/dsse",
        // DSSE signature
        "payload": "eyJfdHlwZSI6ICJodHRwczovL2luLXRvdG...",
        "payloadType": "scai/report/v0.1",
        "signatures": [{ "sig": "MEQCIAZjdOJnQddF14Rpq..." }]
    }
}
\end{lstlisting}

\newpage
\subsection{Report for gcc compilation with SLSA Provenance~\cite{slsa}, DSSE signature~\cite{dsse}}
\begin{lstlisting}[language=json, firstnumber=1]
{
    "subject": {
        "name": "hello-world", 
        "digest": { "sha256": "ced1af6..." },
        "locationURI": "http://example.com/binaries/hello-world"
    },
    "attributes": [{
        "attribute": "WITH_STACK_PROTECTION", 
        "conditions": { "flags": "-fstack-protector*" }
        "evidence": {
            "name": "gcc9.3.0-endorsement", 
            "digest": {"sha256": "abababa..."},
            "locationURI": "http://example.com/scai-reports/",
            "objectType": "scai/report/v0.1"
       	}
    },
    {
        "attribute": "SLSA_PROVENANCE", 
        "evidence": {
            "name": "slsa-provenance-hello-world",
            "uri": "http://example.com/rekor-instance",
            "digest": {"sha256": "0123cdef..."},
            "mediaType": "application/x.dsse+json"
       	}
    }],
    "producer": {
        "uri": "https://example.com/sources/Makefile"
    },
    "signature": {
        "type": "https://github.com/secure-systems-lab/dsse",
        // DSSE signature
        "payload": "eyJfdHlwZSI6ICJodHRwczovL2luLXRvdG...",
        "payloadType": "scai/report/v0.2",
        "signatures": [{ "sig": "MEQCIAZjdOJnQddF14Rpq..." }]
    }
}
\end{lstlisting}

\newpage
\subsection{Report for build with dependency provenance, in-toto Link~\cite{in-toto}, JWS signature~\cite{jws}}
\begin{lstlisting}[language=json, firstnumber=1]
{   
    "subject": {
        "uri": "http://example.com/binaries/some-crypto-app",
        "digest": { "sha256": "efefefe..." },
        "mediaType": "application/x.elf"
    },
    "attributes": [{
        "attribute": "ATTESTED_DEPENDENCY",
        "target": {
            "name": "my-rsa-lib.so", 
            "digest": { "sha256": "ebebebe..." },
            "downloadLocation": "http://example.com/libraries/my-rsa-lib.so"
        } 
        "evidence": {
            "name": "rsa-lib-build-report", 
            "digest": {"sha256": "dcadcad..."},
            "downloadLocation": "http://example.com/scai-reports/rsa-lib-build-report",
            "mediaType": "scai/report/v0.2"
       	}
    },
    {
      	"attribute": "BUILD_LINK", 
        "evidence": {
            "uri": "http://example.com/rekor-instance/crypto-app-intoto-link.json",
            "digest": {"sha256": "4567abcd..."},
            "mediaType": "application/x.dsse+json"
       	}
    }],
    "producer": {
        "uri": "http://example.com/in-toto-builder"
    },
    "signature": {
        "type": "https://datatracker.ietf.org/doc/rfc7515/",
        // JWS signature (flattetend)
        "payload": "eyJfdHlwZSI6ICJodHRwczovL2lu...",
        "protected":"kjbrfkjekrf...",
        "header": { "kid":"deadead..." },
        "signature": "DGkldfhHDG..."
    }
}
\end{lstlisting}
\newpage
\section{Version History}

\begin{center}
\begin{tabular}{ |c|p{12cm}|c| } 
 \hline
 \textbf{Version} & \textbf{Description} & \textbf{Date} \\ 
 \hline
 0.2 & Deprecated Object Reference specification. Added links to in-toto compatible attestation SCAI predicate. Updated all sections. & May 26, 2023 \\
 \hline
 0.1 & Initial version released & Oct 11, 2022 \\ 
 \hline
\end{tabular}
\end{center}

\newpage
\section*{Notices \& Disclaimers}
Intel technologies may require enabled hardware, software or service activation. \\
No product or component can be absolutely secure. \\
Your costs and results may vary. \\
© Intel Corporation. Intel, the Intel logo, and other Intel marks are trademarks of Intel Corporation or its subsidiaries. Other names and brands may be claimed as the property of others.   

\newpage
\bibliographystyle{acm}
\bibliography{references}

\begin{thebibliography}{10}

\bibitem{c2pa}
{\sc {Coalition for Content Provenance and Authenticity (C2PA)}}.
\newblock {C2PA Specifications}.
\newblock \url{https://c2pa.org/specifications/specifications/1.1/index.html}.

\bibitem{dsse}
{\sc {DSSE contributors}}.
\newblock {DSSE: Dead Simple Signing Envelope}.
\newblock \url{https://github.com/secure-systems-lab/dsse}.

\bibitem{solarwinds-hack}
{\sc {FireEye}}.
\newblock {Highly Evasive Attacker Leverages SolarWinds Supply Chain to
  Compromise Multiple Global Victims With SUNBURST Backdoor}.
\newblock FireEye Blogs - Threat Research, Dec 2020.

\bibitem{sgx}
{\sc {Intel Corporation}}.
\newblock {Intel® Software Guard Extensions (Intel® SGX)}.
\newblock \url{https://software.intel.com/en-us/sgx}.

\bibitem{jws}
{\sc {M. Jones, J. Bradley and N. Sakimura}}.
\newblock {JSON Web Signature (JWS)}.
\newblock \url{https://datatracker.ietf.org/doc/rfc7515/}.

\bibitem{oscal}
{\sc {OSCAL NIST Team}}.
\newblock {OSCAL: the Open Security Controls Assessment Language}.
\newblock \url{https://pages.nist.gov/OSCAL/}.

\bibitem{reproducible-builds}
{\sc {Reproducible Builds core team}}.
\newblock Reproducible builds.
\newblock \url{https://reproducible-builds.org/}.

\bibitem{sigstore}
{\sc {Sigstore contributors}}.
\newblock {Sigstore}.
\newblock \url{https://docs.sigstore.dev/}.

\bibitem{slsa}
{\sc {SLSA contributors}}.
\newblock {SLSA Provenance}.
\newblock \url{https://slsa.dev/provenance/v0.2}.

\bibitem{ssdf}
{\sc Souppaya, M., Scarfone, K., and Dodson, D.}
\newblock {Secure Software Development Framework (SSDF) Version 1.1:
  Recommendations for Mitigating the Risk of Software Vulnerabilities}.
\newblock
  \url{https://nvlpubs.nist.gov/nistpubs/SpecialPublications/NIST.SP.800-218.pdf}.

\bibitem{spdx}
{\sc {SPDX contributors}}.
\newblock {SPDX Overview}.
\newblock \url{https://spdx.dev/about/}.

\bibitem{ita}
{\sc {The in-toto/attestation contributors}}.
\newblock {in-toto Attestation Framework}.
\newblock \url{https://github.com/in-toto/attestation}.

\bibitem{in-toto}
{\sc {The Linux Foundation}}.
\newblock {in-toto Specifications}.
\newblock \url{https://in-toto.io/specs/}.

\end{thebibliography}


\end{document}